\documentclass[fleqn]{aastex631}
\usepackage[utf8]{inputenc}

\usepackage{graphicx}	
\usepackage{amsmath}	
\usepackage{amssymb}	
\usepackage{natbib}
\usepackage{comment}
\newcommand{\VEC}[1]{\vec {#1} }
\newcommand{\Grad}{\vec {\nabla} }
\begin{document}

%
\title{
Residual ellipticity driven by intense magnetic stress in the hot neutron star era 
}
\date{\today}
%
\author[0000-0002-4402-3568]{Yasufumi Kojima}
\affiliation{
Department of Physics, Graduate School of Advanced Science and Engineering,\\
Hiroshima University, Higashi-Hiroshima, Hiroshima 739-8526, Japan}

\author[0000-0001-8726-5762]{Akira Dohi}

\affiliation{RIKEN Center for Interdisciplinary Theoretical \& Mathematical Sciences (iTHEMS), RIKEN 2-1 Hirosawa, Wako, Saitama 351-0198, Japan}

\author[0000-0002-9072-4744]{Kotaro Fujisawa}
\affiliation{Division of Liberal Arts, Kogakuin University, Nakanocho 2665-1, Hachioji, Tokyo, 192-0015, Japan}

\author[0000-0002-2498-1937]{Shota Kisaka}
\affiliation{
Department of Physics, Graduate School of Advanced Science and Engineering,\\
Hiroshima University, Higashi-Hiroshima, Hiroshima 739-8526, Japan}
\affiliation{Hiroshima Astrophysical Science Center, Hiroshima University, 1-3-1, Higashi-Hiroshima, Hiroshima 739-8526, Japan}


\begin{abstract}
A neutron star with a nonaxisymmetric mass distribution continuously radiates gravitational waves.
The amplitude depends on the magnitude of the deformation, and is crucial for gravitational-wave astronomy.
We studied an evolutionary effect on the magnetically induced ellipticity in the neutron star crust.
A large deformation at the hot era before solidification is assumed to be produced magneto-hydrodynamically, and the fate at later time follows owing to the decrease in the force associated with the decay and redistribution of the current.
Instead of simulating the evolution of deformation with the magnetic field over a long timescale,
we estimate the ellipticity that remains after the formation of the solid crust by comparing two equilibrium models before and after solidification.
The relic ellipticity is $\sim 10^{-8} (B_0/10^{13}{\rm{G}})^2$ in the magnitude for dipole field strength $B_{0}$ at the surface.
This value was primarily determined by the magnetic field at the later time, with a weak dependence on the initial large deformation. 
The memory of the presolidification state was retained, but was small because 
the elastic force contributes little to the equilibrium.
\end{abstract}

\section{Introduction}

A neutron star (NS) is expected to be almost spherical, and the degree of nonspherical deformation is very small owing to strong gravity.
However, a rotating NS with nonaxially symmetric mass distribution continuously emits gravitational waves (GWs) at fixed frequency.
GW astronomy has been explored such an evidence~\cite[e.g.,]
[for a review of continuous GW]{2023LRR....26....3R,2023APh...15302880W}, but the signal is not yet detected.
According to current observation of continuous GW~\citep[][]{2026ApJ..1005..221A}, the upper limits of quadrupole deformation fall below the spindown limit for 20 known pulsars.
Recently, a stringent limit of $\epsilon < 10^{-8}$ is reported for the millisecond pulsar J0435+3233~\citep{2026arXiv260718219M}, where $\epsilon$ is the ellipticity characterizing quadrupole deformation.
However, the constraint is significantly relaxed if the pulsar is located in a hierarchical triple system~\citep{2026arXiv260727462F,2026arXiv260801227Y}.
Further developments to detect the GW signal are promising~\citep[see][for a review]{Corsi+24}.
Detection in the future will provide useful information on the mechanism of deformation and interior structure. 
As a rotating fluid star is symmetric with respect to the spin axis, additional mechanisms are necessary to account for the nonaxisymmetric deformation relevant to GW radiation.
We do not consider the nonaxisymmetric oscillation of the r-mode in a rapid rotator~\citep{1998PhRvD..58h4020O,1999ApJ...516..307A}.
A possible candidate for deformation is the intense magnetic field inside neutron stars because the direction of magnetic deformation is independent of the stellar spin.
The ellipticity $\epsilon$ is estimated by the ratio of the magnetic to gravitational energy:
 $\epsilon \sim 10^{-10} ( B/10^{13}{\rm{G}} )^{2}$, where $B$ is the average magnetic-field strength inside the star.
For stars with type-II superconductor cores, the ellipticity is given by $\epsilon  \sim 3\times 10^{-8}$ $( H_{c}/10^{15}{\rm{G}} )(B/10^{13}{\rm{G}})$~\citep{2012MNRAS.419..732L,2014MNRAS.437..424L},
where $H_{c}$ is the lower critical field strength required for superconductivity.
The geometry and strength of the magnetic field are crucial for obtaining more definite results beyond order-of-magnitude estimates
(\cite{1996A&A...312..675B,1999A&A...352..211K,2021EPJA...57..249C} 
for detailed modeling using various approaches).
A solid crust is another possible origin of ellipticity $\epsilon$.
Such a crust can sustain small irregularities due to shear stress and will be fractured when the deformation force is excessively strong. Thus, the deformation can be determined using the ratio of the elastic force to gravity.
The maximum ellipticity of a solid crust is estimated using the shear modulus $\mu$ and maximum strain $\sigma_{\rm{max}}$:
$\epsilon \sim $ $(\Delta M/M)\times$ 
$(\mu \sigma_{\rm{max}}/\Delta r)/( GM\rho_{\rm{crust}}/R^2) \sim$
$10^{-6} (\sigma_{\rm{max}}/0.1)$, where $\Delta M$ is the mass, $\Delta r$ is the thickness, and
$\rho_{\rm{crust}}$ is the mass density of the crust.
See e.g.,~\cite{2000MNRAS.319..902U,2006MNRAS.373.1423H,2013PhRvD..88d4004J} for detailed calculations.
The actual ellipticity is smaller and dependent on the explicit mechanism.
As a deforming mechanism, \citet{2000MNRAS.319..902U} discussed the temperature anisotropies in the accreted crust, for which the resultant ellipticity is $\epsilon = 2.8 \times 10^{-6}$.
See also the recent detailed calculations of the thermal mountain
\citep{2023MNRAS.522..226H,2025MNRAS.540.2349J}, and ellipticity based on anisotropic shear intrinsic to the NS~\citep{2024PhRvD.110d4016M}.
Recently, a plausible value of $\epsilon$ in an isolated pulsar was studied in an evolutionary scenario~\citep{
2021MNRAS.500.5570G,2022MNRAS.517.5610M,2026MNRAS.546ag141K}.
They considered how much quadrupole deformation remains through the solid crust formation phase of a proto-NS, assuming that a large deformation was initially driven in a hot era at birth.
The results of our previous work~\citep{2026MNRAS.546ag141K} revealed that even if the deformation is assumed to be large in the hot era immediately after birth, it significantly shrinks upon thermal cooling.
The reduction factor is typically $10^{-3}$, and 
$\epsilon \sim 10^{-5}\zeta$ remains throughout the crustal solidification at $\sim 10^2-10^4$ s of age, where $\zeta$ is a dimensionless parameter characterizing the magnitude of the deforming force $0< \zeta < 1$.
The perturbation amplitude is proportional to $\zeta$, and the maximum $\zeta =1$ corresponds to the limit of the linear theory.  
Studies have shown that a small amount of deformation is imprinted onto the solid crust.
However, the driving force used was an abstract one, and the magnitude of the deformation force was unclear.
The residual value significantly depends on the forces~\citep{2021MNRAS.500.5570G,2022MNRAS.517.5610M}; therefore, further considerations are necessary to obtain more reliable values. 
Unlike our previous work~\citep{2026MNRAS.546ag141K}, in which a mathematical force with parametrized strength $\zeta$ is used,
we here consider specific magnetic field configurations and calculate the residue of magnetically induced ellipticity, which is expressed by observable quantities.
The strength of the magnetic field as well as the geometry at the hot era of proto-NS are largely uncertain, but is likely to be larger than that observed in normal radio pulsars with $B \sim 10^{13}$G (see also \cite{2026arXiv260116159P}).
It may reach $B \sim 10^{15}$G~\citep{2009A&A...499..557R} assuming the equipartition of magnetic with thermal energies
above the solidification temperature $T_c \sim 6.4\times 10^{9}$K.
The equilibrium at the hot era is likely to be nonbarotropic, i.e., the surface for constant pressure differs from that for density.
A much larger family of magnetic-field equilibria is possible~\citep{2009A&A...499..557R,2011MNRAS.417.2288M,2013MNRAS.433.2445A,2016MNRAS.463.2542G}.
As a star cools, the structure tends to be barotropic with a loss of thermal pressure.
Does the large deformation before solidification significantly affect the equilibrium with the elastic force?
How much deformation in the past is imprinted on the crust at present?
The evolutionary approach is essential to explore crustal deformation, which is relevant to continuous gravitational wave searches.
This evolutionary notion differs from some studies of magnetoelastic deformation~\cite[e.g.,]
[]{2021MNRAS.506.3936K,2022MNRAS.511..480K}, in which elastic deformation was calculated against the magnetic force.
The magneto-thermal evolution through the solidification era is necessary to obtain the relic deformation.
After the formation of a solid crust, the magnetic field evolution is governed by the Hall–Joule equation~\citep[e.g.,][for a review]{2026LRCA...12....5P}.
Numerical simulations have revealed complex magnetic-field structures with significant higher-order multipole components arising from nonlinear mode coupling 
\cite[e.g.,][for recent 3D models]{2022ApJ...936...99D,2025A&A...694A..39D,2026ApJ..1000..291I}.
The resulting field configurations depend on the strength and geometry of the initial magnetic field.
The simulations however do not cover the regime relevant to our concern.
The concern is the long-term evolution of the magnetic field, and the initial condition corresponds to that for lower temperature ($T \sim 10^8$K) because the hot era is too short owing to intense neutrino cooling. The deformation resulting from the force balance was not considered.
The quadrupole deformations associated with the magnetic Hall--Joule evolution with an initial symmetric state were estimated with a few exceptions~\citep{2016MNRAS.459.3407S,2021MNRAS.502.2097K}.
It remains uncertain whether the approach considered in previous works can be extended to the dynamics associated with the fluid-to-solid phase transition.
In this paper, we did not follow the time evolution of deformation in which a combined system of hydrodynamics, magnetic field, and thermal state must be simultaneously solved.
Such a numerical calculation on a long timescale is extremely difficult, and is beyond the scope of this study.
We adopt the idea of the ``evolutionary'' effect considered in previous studies~\citep{
2021MNRAS.500.5570G,2022MNRAS.517.5610M,2026MNRAS.546ag141K}, that is, elastic deformation remaining at a later time is calculated by comparing static equilibria at two different times, assuming that the magnetic field changes during the interval.
This quasi-static approach is justified when the magnetic field evolution timescale by the Hall drift and Ohmic decay is long compared to the hydrodynamic timescale, allowing the crust to maintain mechanical equilibrium at each stage. 
By approximation, we estimate the effect of large deformations before solidification on the force balance incorporated with elastic force at the later time.
The remainder of this paper is organized as follows. 
Section 2 describes the mathematical formulation and assumptions.
We solve the crustal part in the mass density range $1.4 \times 10^{14} - 4 \times 10^{11} $gcm$^{-3}$, and calculate the $l=2$ part of the deformation for the ellipticity $\epsilon$.
The magnetic field models driving the deformation are presented in Section 3.
In Section 4, the numerical results of the residual ellipticity are presented and compared with those for different models and physical situations.
Finally, the conclusions are presented in Section 5.

\section{Mathematical formulation}
We consider an equilibrium slightly deformed from a spherical star 
due to the Lorentz force ${\VEC {f}} = {\VEC {j}}\times {\VEC {B}}/c$.
The dominant forces are pressure and gravity forces. The static equilibrium in Newton gravity is approximated by
\begin{equation}
\Grad p_{0} + \rho_{0}\Grad \Phi_{0} =0.
\label{Forcebalance0.eqn}
\end{equation}
Our analysis neglects the stellar rotation, which may be added as a centrifugal term in the gravitational potential.
The centrifugal deformation was axially symmetric and irrelevant to the GW radiation.
A complete treatment of a rotating inclined dipole magnetic field would require solving for the combined magneto-rotational equilibrium, which is beyond the scope of this study.
The magnitude of the Lorentz force is so small that it can be treated as a perturbation in the background equilibrium given by Eq.~(\ref{Forcebalance0.eqn}):
\begin{equation}
\Grad \delta p +\delta \rho \Grad \Phi_{0} +\rho_{0} \Grad \delta \Phi = {\VEC{f}},
\end{equation}
where $\delta$ denotes the Eulerian variation.
We consider a spherical core and the deformation of the crustal region, which denotes the radial region $r_c\le r \le R$, regardless of whether the solid crust forms. 
We use a spherical coordinate $(r, \theta, \phi)$ and limit our consideration to the axially symmetric configuration $\partial_{\phi}=0$.
The magnetic dipole axis may be chosen as the symmetric axis because we neglect the stellar rotation. 

  \subsection{Crust as a background model}
We consider the crust of neutron stars, with mass densities ranging from $\rho_{c}$ $=1.4\times 10^{14}$ g cm $^{-3}$ to the neutron-drip density $\rho_{1}=4\times 10^{11}$ g cm$^{-3}$~\citep{2026MNRAS.546ag141K}.
The inner radius was $r_{c}$ at the core--crust interface, and the outer radius was chosen as the stellar surface $R$, because
we ignore the extremely thin outer crust corresponding to the low density, which contributes negligibly to the ellipticity.
The thickness was assumed to be $\Delta r/R=(R-r_{c})/R=0.1$.
Equation~(\ref{Forcebalance0.eqn}) is approximated as
\begin{equation}
    p_{0} ^{\prime} = -\frac{GM_{r} \rho_{0} }{r^2}
  \approx    -\rho_{0} g_s\left(\frac{R}{r}\right)^2,
\label{hydroequil.eqn} 
\end{equation}
where $M_{r}$ is the mass within a radius $r$.
The crustal mass is extremely small compared with the total mass $M$; hence, $M_{r}$ can be approximated as $M$.
The gravitational acceleration $g_s$ at the surface is chosen as $g_s = GM/R^2 \approx 1.5 \times 10^{14}$ cm s$^{-2}$ for $M \approx 1.5 M_{\odot}$ and $R\approx 12$ km.
The solution of Eq.~(\ref{hydroequil.eqn}) is reasonably approximated in an analytic form with a polytropic index $n=2$. 
\begin{align}
    \rho_{0} & =\rho_{c}\left[1+\alpha(r^{-1}-r_c ^{-1} )\right]^2,
   \label{polytropic1.eqn} 
   \\
     p_{0} & =p_{c}
     \left[1+\alpha(r^{-1}-r_c ^{-1} )\right]^3,
   \label{polytropic2.eqn} 
\end{align}
where $\alpha = (1-(\rho_{1}/\rho_{c})^{1/2})r_c R/\Delta r$. The pressure $p_c$ at $r=r_c$ is given by
 $p_c  = \rho_c g_s R^2 /(3\alpha)$  $\approx 9.8 \times 10^{32}$ dyn cm$^{-2}$, and radially decreases 
$p_{0}(R) =p_{c}(\rho_{1}/\rho_{c})^{3/2}\approx 1.5 \times 10^{-4} p_{c}$
$\approx 1.5 \times 10^{29} $ dyn cm$^{-2}$.
The crustal mass $\Delta M$ is determined to be extremely small at
$\Delta M = 4\pi \int \rho_{0} r^2 dr $ $\approx 3 \times 10^{-2} M$, integrating it from $r_c$ to $R$.
We introduce a sound velocity $c_s$ for the spatial profile of Eqs.~(\ref{polytropic1.eqn})--(\ref{polytropic2.eqn}):
\begin{equation}
    c_s = \left( \frac{p_{0} ^\prime}{\rho_{0} ^\prime}
    \right)^{1/2} .
    \label{def_cs.eqn}
\end{equation}
This ranges from $c_s =3.2 \times 10^{9}$cm s$^{-1}$ at $r=r_c$ to $7.5 \times 10^{8}$cm s$^{-1}$ at $R$.
The shear modulus $\mu$ in the solid crust may be approximated as a linear function of density, which is fitted to the results of a detailed calculation~\citep{2008LRR....11...10C}.
Thus, $\mu$ is given by a radial function, approximated in terms of the crustal spatial density profile:
 \begin{equation}
    {\mu} = \mu_{c}
    \left[1+\alpha(r^{-1}-r_c ^{-1} )\right]^2,
 \end{equation}
where $\mu_{c}=10^{30}~{\rm erg~cm}^{-3}$ at the core--crust interface.

\subsection{Lorentz force}
Our consideration is limited to the axially symmetric magnetic field.
The poloidal and toroidal components of the magnetic field are expressed in terms of two functions $\Psi$ and $S$ as
\begin{equation}
\VEC{B}=\VEC{\nabla}\times \left(\frac{\Psi}{\varpi}\VEC{e}_{\phi}\right)
+\frac{S}{\varpi}\VEC{e}_{\phi},
\end{equation}
where $\varpi=r\sin\theta$, and
$\VEC{e}_{\phi}$ is the azimuthal unit vector in spherical coordinate $(r,\theta, \phi)$.
An electric current is calculated by the Amp{\'e}re--Biot--Savart law $4\pi c^{-1}{\VEC{j}}=\Grad \times \VEC{B}$.
The azimuthal component of the Lorentz force $\VEC{f}= c^{-1}{\VEC {j}}\times {\VEC {B}}$ should vanish for an axially symmetric model so that the function $S$ is constrained as $S=S(\Psi)$.
The poloidal component of the acceleration ${\VEC {a}}\equiv \rho_{0} ^{-1} \VEC{f}$ is written as
\begin{equation}
{\VEC {a}} =J \Grad \Psi ,
\label{avectJ.eqn}
\end{equation} 
where
\begin{equation}  
 J \equiv
 \frac{1}{4\pi \varpi^2 \rho_{0}}\left(\frac{4\pi\varpi}{c}j_{\phi} - 
 SS^{\prime} \right).
\label{DefJ3.eqn}
\end{equation} 
The magnetic flux function $\Psi$ is obtained by solving 
\begin{equation}
\left(\frac{\partial ^2}{\partial r^2}+
\frac{\sin \theta }{r^2}\frac{\partial}{\partial \theta}
\frac{1}{\sin \theta }\frac{\partial}{\partial \theta}
\right)\Psi = -\frac{4\pi \varpi }{c} j_{\phi}.
\label{Psi2D.eqn}
\end{equation}
For a barotropic equilibrium, in which $\Grad \times  \VEC{a} =0$, the function $J$ is constrained as $J=J(\Psi)$.
Thus, the magnetic field is determined once the functions $S(\Psi)$ and $J(\Psi)$ are given. 
However, nonbarotropic models are free from the constraint to $J$; 
therefore, various magnetic fields are possible~\citep{2009A&A...499..557R,2011MNRAS.417.2288M,2013MNRAS.433.2445A,2016MNRAS.463.2542G}.
We assume a linear form $K=k_0\Psi$ ($J=k_0$) and $S=\kappa \Psi$ to obtain a simple model of the dipolar magnetic field.
By expressing $\Psi = g_{1}(r) \sin^2 \theta$ and $j_{\phi} = j_{1}(r) \sin \theta$, Eq.~(\ref{Psi2D.eqn}) is reduced to
\begin{equation}
\frac{d^2 g_{1} }{dr^2} -\frac{2 g_{1}}{r^2} =-\frac{4\pi r}{c}j_{1} .
\label{diffeqg1.eqn}
\end{equation}
We choose $J$ in Eq.~(\ref{DefJ3.eqn}) as a radial function $s_1(r)$. The radial function $j_1$ is expressed as
\begin{equation}
\frac{4\pi r}{c}j_{1} = 4\pi \rho_{0} r^2  s_{1} +\kappa ^2 g_{1}.
\label{sourcej1.eqn}
\end{equation}
Our choice of $s_1$ is discussed in the next subsection. Note that a case of $s_1=k_0$ represents a barotropic model.
The acceleration due to the Lorentz force for the dipolar magnetic field induces perturbations of the monopole ($l=0$) and quadrupole ($l=2$):
\begin{equation}
{\VEC {a}}= s_{1} \Grad (g_{1} \sin^2 \theta ) =-\frac{2}{3} s_{1}\Grad g_{1}(P_2 -P_0),
\label{EMForcedipole.eqn}
\end{equation}
where $P_{l}(\cos\theta)$ is the Legendre polynomial of order $l$, and $P_0=1$, $P_2=(3\cos^2 \theta-1)/2$ are used.
The angular parts of perturbations are expanded by $P_{0}$ and $P_2$, e.g.,  $\delta \rho$ $= \delta \rho_{0} (r) + \delta \rho_{2} (r) P_2(\cos\theta)$.
The force-balance equations are decoupled by different $P_{l}$. We consider the $l=2$ part only to calculate the quadrupole ellipticity.

\subsection{Deformation by fluid description}
\label{sec2.3}
We consider small deformations in the fluid from a spherical symmetric background model. The perturbations are denoted with a subscript $_a$ as $\delta p_{a}, \delta \rho_{a} $, and $\delta \Phi_{a}$.
The force balance is determined by
\begin{equation}
\Grad \delta p_{a} +\delta \rho_{a} \Grad \Phi_{0} +\rho_{0} \Grad \delta \Phi_{a} = 
{\VEC{f}}_a.
\label{Forcebalance1a.eqn}
\end{equation}
This equation can be written equivalently as
\begin{equation}
\Grad \left( \frac{\delta p_{a} }{\rho_{0}} + \delta \Phi _{a}
\right)  + \frac{\rho_0 ^{\prime}}{\rho_{0} ^2}
Q \VEC{e}_{r}= \frac{1}{\rho_{0}}\VEC{f}_a,
\label{balanceBF.eqn}
\end{equation}
where
$\VEC{e}_{r}$ is a radial unit vector, and $Q$ is defined for later convenience as
\begin{equation}
Q \equiv \delta p_{a} -\frac{p_0^{\prime}}{\rho_0 ^{\prime}} \delta \rho_a=
\delta p_{a} - c_s ^2 \delta \rho_{a}  ,
\label{DefQterm.eqn}
\end{equation}
where $c_s ^2 =p_0 ^{\prime}/\rho_0 ^{\prime}$ (Eq.~(\ref{def_cs.eqn})) is used.
Equation~(\ref{balanceBF.eqn}) shows that the condition $Q=0$ is incompatible for rotational force $\Grad \times (\rho_{0}^{-1} \VEC{f}) \ne 0$.
The variation in pressure is generally associated with that in the density and other parameters.
We consider a simple case that depends on two parameters: $p=p(\rho,s)$, where $s$ is the entropy per baryon. The variation can be written as
\begin{equation}
\Delta p = \left(\frac{\partial p}{\partial \rho} \right)_s \Delta \rho+ 
\left(\frac{\partial p}{\partial s} \right)_\rho \Delta s ,
\end{equation}
where $\Delta$ denotes the Lagrangian change with the displacement vector ${\VEC{\xi}}$, which connects fluid elements in the background equilibrium with corresponding ones in the perturbed configuration.
The relation between the Eulerian and Lagrangian changes is given by, e.g., $\Delta \rho = \delta \rho +{\VEC{\xi}} \cdot \Grad \rho_{0}$.
Using an adiabatic index, $(\partial p/\partial \rho)_s =\Gamma p/\rho $, Eq.~(\ref{DefQterm.eqn}) is reduced to
\begin{equation}
Q=  \left( 
\frac{\Gamma p}{\rho} -\frac{p_0 ^{\prime}}{\rho _0 ^{\prime}}
\right)\Delta \rho + \left(\frac{\partial p}{\partial s} \right)_\rho \Delta s .
\label{DefQ.eqn}
\end{equation}

The condition $Q \ne 0$ arises from the stratification of the fluid in the first term of Eq.~(\ref{DefQ.eqn}) and/or
the entropy distribution different from the background model in the second term. 
Instead of a detailed microscopic model for $Q \ne 0$,  
we used a constrained form of the force balance with the assumed force $\VEC{f}$, as discussed below.
As the thermodynamic relation between $\delta p_{a}$ and $\delta \rho_{a}$, we use $\delta p_{a} = c_s ^2\delta \rho_{a} + Q$, i.e., Eq.~(\ref{DefQterm.eqn}).
This point differs from the method in other studies~\citep{2021MNRAS.500.5570G,2022MNRAS.517.5610M}.
In taking the curl of Eq.~(\ref{balanceBF.eqn}) by expressing the acceleration $\rho_{0}^{-1} \VEC{f}_a = J\Grad{\Psi}_a$ due to the Lorentz force (\ref{avectJ.eqn}), we have
\begin{equation}
  {\VEC{e}}_{r} \times \Grad
  \left(\frac{\rho_0 ^{\prime} Q}{\rho_{0} ^2}
  \right)  +\Grad  J\times \Grad \Psi_{a} =0.
  \label{relationMgJP.eqn} 
\end{equation}
The first term vanishes for the barotropic case, such that (\ref{relationMgJP.eqn}) is satisfied for $J=K^{\prime}(\Psi_{a})$,
where $K$ is an arbitrary function of $\Psi_{a}$.
The acceleration in Eq.~(\ref{avectJ.eqn}) is given by $\VEC{a}=\Grad K$.
The general solution of $J$ in Eq.~(\ref{relationMgJP.eqn}) is difficult to obtain. 
One possible form is given by an arbitrary radial function $s_1(r)$:
\begin{equation}
 J=s_{1} (r) + K^{\prime}(\Psi_{a}),
\label{JbySandK.eqn}
\end{equation}
and the function $Q$ is expressed as
\begin{equation}
 Q=-\frac{\rho_0 ^{2}  s_{1} ^\prime}{\rho_0 ^\prime} \Psi_{a} .
  \label{NonB_q.eqn}
\end{equation}
We limit $K$ to a linear form; $K=k_0 \Psi_{a}$,$J=k_0$, where $k_0$ is a constant.
Such a simple case is obtained by $s_1= k_0$, such that Eq.~(\ref{JbySandK.eqn}) may be simply written as $J=s_1$.
The acceleration (\ref{avectJ.eqn}) is given by $\VEC{a}= s_1 \Grad \Psi_{a}$.
The deviation from the spherical symmetric state was obtained by integrating Eq.~(\ref{balanceBF.eqn}) with Eqs.~(\ref{JbySandK.eqn})--(\ref{NonB_q.eqn}) as
\begin{equation}
 \frac{\delta p_{a} }{\rho_{0}} + \delta \Phi_{a} =   s_1 \Psi_{a},
\label{integral.eqn}
\end{equation}
where an integration constant may be involved in the above equation but is freely chosen by the constant with $s_1$.
The variations in pressure and density were solved with Eqs.~(\ref{DefQterm.eqn}), (\ref{NonB_q.eqn}), and (\ref{integral.eqn}) as follows:
\begin{align}
\delta p_{a} &= \rho_0 ( -\delta \Phi_{a} +  s_1 \Psi_{a}),
\label{fld_pr_phi.eqn}
\\
 \delta \rho_{a} &= \frac{\rho_0 }{c_s ^2} 
   \left( -\delta \Phi_{a} 
   +\frac{(\rho_{0} s_{1})^{\prime}}{\rho_0 ^{\prime}} \Psi_{a}  \right).   
 \label{fld_dns_phi.eqn}
\end{align}
The gravitational perturbation $\delta \Phi_{a}$ is determined by
\begin{equation}
 \nabla ^2 \delta \Phi_{a} = 4 \pi G_{\rm{N}} \delta \rho_{a} .
\label{Poisson.eqn}
\end{equation} 
The differential equation (\ref{Poisson.eqn}) is solved with the boundary conditions for the deformed crust~\citep[see][]{2026MNRAS.546ag141K}.
Thus, we determine the quadrupole deformation by the radial functions $\delta \rho_{a2}$, $\delta p_{a2}$, and $ \delta \Phi_{a2}$,
which are split by the Legendre polynomial $P_{l}$ in Eqs.~(\ref{fld_pr_phi.eqn})--(\ref{Poisson.eqn}).

\subsection{Deformation by fluid-elastic description}
\label{sec2.4}
In a solid crust, an elastic force ${\VEC{h}}$ is involved in the force balance.
Small variations from a spherical background model are expressed with the subscript $_b$ (e.g., $\delta p_{b}$). Equation~(\ref{Forcebalance1a.eqn}) for a static equilibrium, denoted as B, is modified as
\begin{equation}
\Grad \delta p_{b} +\delta \rho_{b} \Grad \Phi_{0} +\rho_{0} \Grad \delta \Phi_{b} + \VEC{h} = 
{\VEC{f}}_b,
\label{ElasticEquil.eqn}
\end{equation}
where components of the elastic force are explicitly given by
\begin{align}
   & h^{i} = -\nabla_{j} ( 2\mu\sigma^{ij} ),
    \\
& \sigma_{ij} =\frac{1}{2}
\left( \nabla_{i} \xi_{j} +\nabla_{j} \xi_{i}  \right)
-\frac{1}{3} g_{ij} \nabla_{k} \xi^{k} .
\end{align}
We consider the polar displacement $\VEC{\xi}=(\xi_r, \xi_\theta, 0)$, which is relevant to the variation in density. 
\begin{equation}
\delta \rho_{b} = -{\Grad}\cdot (\rho_{0} {\VEC{\xi}}) .
\label{densitybyxi.eqn}
\end{equation}
This equation represents the mass conservation associated with ${\VEC{\xi}}$.
The equilibrium state B is assumed to be related to the adiabatic change from the background model; hence, the thermodynamic relation between $\delta p_{b}$ and $\delta \rho_{b}$ is given by 
\begin{equation}
  \delta p_{b} = c_s ^2 \delta  \rho_{b}. 
   \label{thermBstate.eqn}
\end{equation}
The gravitational perturbation $\delta \Phi_{b} $ is determined by
\begin{equation}
\nabla ^2 \delta \Phi_{b} =4 \pi G_{\rm {N}} \delta \rho_{b} .
\label{PossionPT2.eqn}
\end{equation}
The displacement $\VEC{\xi}=0$ corresponds to the spherical model without $\VEC{f}_b$. In other words, a set of equations (\ref{ElasticEquil.eqn})--(\ref{PossionPT2.eqn}) describes an elastic deformation imposed by the Lorentz force $\VEC{f}_b$. Other hydrodynamic perturbations were also induced by $\VEC{f}_b$.

\subsection{Deformation in evolutionary scenario}
\label{sec2.5}
We consider the evolution of the deformation produced by the force $\VEC{f}_a$ in the hot era.
  The force gradually changes to $\VEC{f}_b$ with decreasing thermal energy. A new equilibrium state is reached incorporated with an elastic force after the solid crust forms below the solidification temperature $T < T_{c}$.
We assume that matter in hot state A is nonbarotropic, and barotropic in cold state B.
The static structure at each state can be obtained once the magnetic field is given.
We do not simulate the evolution of the magnetic field over a long timescale, but assume the field at states A and B.
It is difficult to determine whether the evolutionary path from A to B is realistic. 
However, the model calculation in the quasi-static approximation demonstrates  
the extent to which the large magnetically driven deformation in the hot presolidification-era affects the final elastic equilibrium.
The deviation in density from the background model changes from $\delta \rho_{a}$ to $\delta \rho_{b} \equiv \delta \rho_{a} + \delta \rho_{+}$.
Similarly, we also write $\delta p_{b} =\delta p_{a} + \delta p_{+}$ and
 $\delta \Phi_{b} =\delta \Phi_{a} + \delta \Phi_{+}$.
The perturbation equation describing state B is the same as Eq.~(\ref{ElasticEquil.eqn}).
However, the displacement $\VEC{\xi}=0$ corresponds to state A.
The variation in density is described as
\begin{equation}
\delta \rho_{+} = -{\VEC{\nabla}}\cdot (\rho_{*} {\VEC{\xi}}) 
\approx -{\VEC{\nabla}}\cdot (\rho_{0} {\VEC{\xi}}) ,
\label{densplusbyxi.eqn}
\end{equation}
where $\rho_{*}$ is the mass density in state A, and is only different from $\rho_{0}$ by a small quantity in the first order.
By neglecting small quantities in the second order, we approximated $\rho_{*}$ as $\rho_{0}$ in Eq.~(\ref{densplusbyxi.eqn}).
By subtracting Eq.~(\ref{Forcebalance1a.eqn}) from Eq.~(\ref{ElasticEquil.eqn}), we obtain 
\begin{equation}
\Grad \delta p_{+} +\delta \rho_{+} \Grad \Phi_{0} +\rho_{0} \Grad \delta \Phi_{+} + \VEC{h} = 
{\VEC{f}}_b - {\VEC{f}}_a.
\label{Forcebalance2.eqn}
\end{equation}
In previous studies ~\citep{2021MNRAS.500.5570G,2022MNRAS.517.5610M,2026MNRAS.546ag141K}, $|{\VEC{f}}_b| \ll  |{\VEC{f}}_a |$ and the right-hand side in Eq.~(\ref{Forcebalance2.eqn}) was approximated as $-\VEC{f}_{a}$.
In this study, ${\VEC{f}}_b$ is taken into account, although the magnitude is smaller than that of ${\VEC{f}}_a$.
The adiabatic relation between the density and pressure variations from the background model is assumed to be satisfied.
The relation between $\delta p_{b}$ and $\delta \rho_{b}$ is given by Eq.~(\ref{thermBstate.eqn}).
The thermodynamic relation between $\delta p_{+}$ and $\delta \rho_{+}$ is  given by
\begin{equation}
  \delta p_{+} = c_s ^2 \delta  \rho_{+} -Q. 
   \label{p2withq2pls.eqn}
\end{equation}
The change in the density $\delta \Phi_{+}$ is related to the gravitational potential $\delta \rho_{+}$ as
\begin{equation}
\nabla ^2 \delta \Phi_{+} =4 \pi G_{\rm {N}} \delta \rho_{+} .
\label{PossionPT3.eqn}
\end{equation}
A set of equations for the quantity with the subscript $_{+}$ is given by Eqs.~(\ref{densplusbyxi.eqn})--(\ref{PossionPT3.eqn}), and similar to those for the subscript $_{b}$.
However, they are different in the force and thermodynamic relation.
The direction of ${\VEC{\xi}}$ is also different.
The relic deformation is a sum, e.g.,
$\delta \rho_{b}=\delta \rho_{a} +\delta \rho_{+}$, of the initial deformation and the variation that arises from ${\VEC{f}}_a$ to ${\VEC{f}}_b$.
We also expand these perturbation equations by the Legendre polynomial $P_{l}$, and consider only the $l=2$ part.
The numerical method and boundary conditions are provided in~\cite{2026MNRAS.546ag141K}.

\subsection{Quadrupole ellipticity}
\label{sec2.6}
The ellipticity $\epsilon$ characterizing the quadrupole deformation
is a dimensionless quantity given by
\begin{equation}
     \epsilon  
     = \frac{I_{xx}-I_{yy}}{I_{zz}}
          = \frac{2\pi}{MR^2}\int_{r_c} ^{R} \delta \rho_{2} r^4 dr,
\label{def_ellp.eqn}
\end{equation}
where the denominator in Eq.~(\ref{def_ellp.eqn}) is approximated using $2MR^2/5$ as the uniform density of a spherical star with mass $M$ and radius $R$. However, the deformed crust is relevant to the numerator.

We introduce different types of ellipticities: $\epsilon_{a}$, $\epsilon_{*}$, $\epsilon_{z}$, and $\epsilon_{s}$, depending on their deformation process. 
First, $\epsilon_{a}$ characterizes the deformation in the fluid system. 
This corresponds to the value produced at the hot era, or a case of inefficient solidity. The equilibrium structure at the hot state is assumed to be nonbarotropic. The calculation of the deformation is shown in Section~\ref{sec2.3}, and $\epsilon_{a}$ is calculated for $\delta \rho_{a2}$.
This value may be irrelevant to the current observation because the thermal structure and magnetic fields are those of the past and have already changed.
Second, $\epsilon_{*}$ indicates that the relic ellipticity remained owing to the transition of the magnetic force. The magnetic force in the later state is given by ${\VEC{f}}_b$, for which the magnetic field is assumed to be barotropic. The calculation of the deformation is explained in Section~\ref{sec2.5}, and $\epsilon_{*}$ is calculated for $\delta \rho_{b2}$$=\delta \rho_{a2}+\delta \rho_{+2}$.
Third, $\epsilon_{z}$ is also a relic deformation, and is obtained by $\epsilon_{*}$ in the limiting case of $|{\VEC{f}}_b| \to 0$.
Finally, $\epsilon_{s}$ is the ellipticity sustained by the magnetic force ${\VEC{f}}_b$. The displacement ${\VEC{\xi}}=0$ corresponds to the background spherical model in which other fluid perturbations vanish.
The value $\epsilon_{s}$, whose method is given in Section~\ref{sec2.4}, is relevant to the magneto-elastic deformation after the solid crust forms.
%

\section{Magnetic Field Models}

\begin{table}[]
\caption{Model parameters and characteristics.
    $a$, $b$, $w$, $\kappa$, and $k_0$ are constant. 
    Negative value $B_s$ in Model IV--VI denotes the opposite direction of the magnetic field.}
\centering
    \begin{tabular}{c|c|c|c|l}
        Model & $B_s/B_0$ &  $E_m/E_0$ &  $J_{\phi} /J_{0}$& current density $j_1 $   \\
         \hline
%
     I &  1.65 & 1.98 & 1.0&
     $\rho_0 r(a+b(r-r_c)^3(r-R))$  \\
    II & 3.29 & 10.0& 2.60&
     $\rho_0 r(a+b(r-r_c)^2(r-R)^2)$\\
    III & 6.65 & 50.0& 6.71&
     $\rho_0 r(a+b(r-r_c)(r-R)^3)$ \\
    IV &  $-0.44$ & 10.0& 2.0&
    $a(2+(wr)^2)r^{-3}\sin(w(r-r_c))$\\
    V & $-1.48$ & 10.0& 1.55&
    $a(2+(wr)^2)r^{-7}\sin(w(r-r_c))$\\
    VI & $-1.25$ & 10.0& 2.42&
    Model IV  and  $\kappa^2\ne 0$\\
\hline
B & $1$ & 1& 1&    $k_0 \rho_0 r$ \\
\hline
    \end{tabular}
\label{table:model}
\end{table}

\begin{figure}\begin{center}
    \includegraphics[width=0.75\columnwidth]{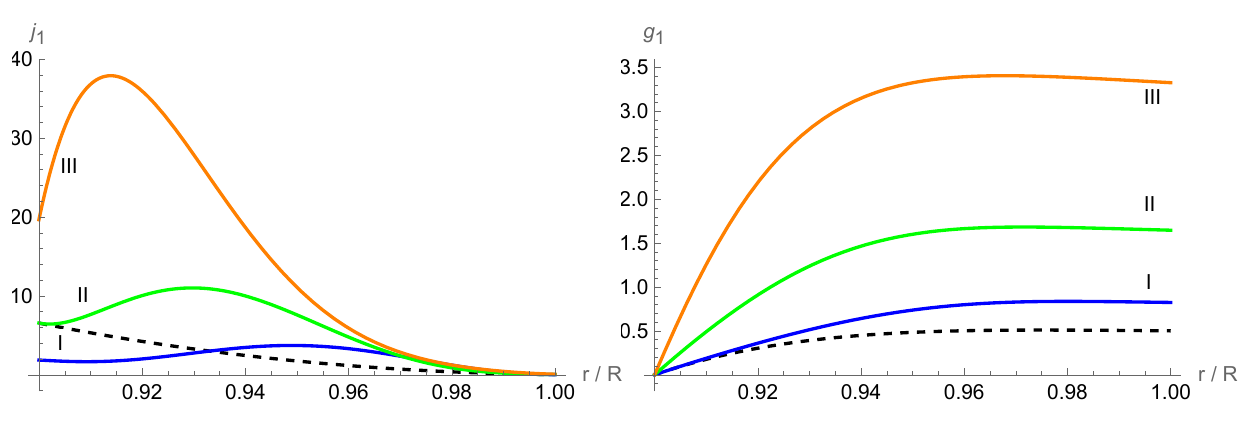}
    \includegraphics[width=0.75\columnwidth]{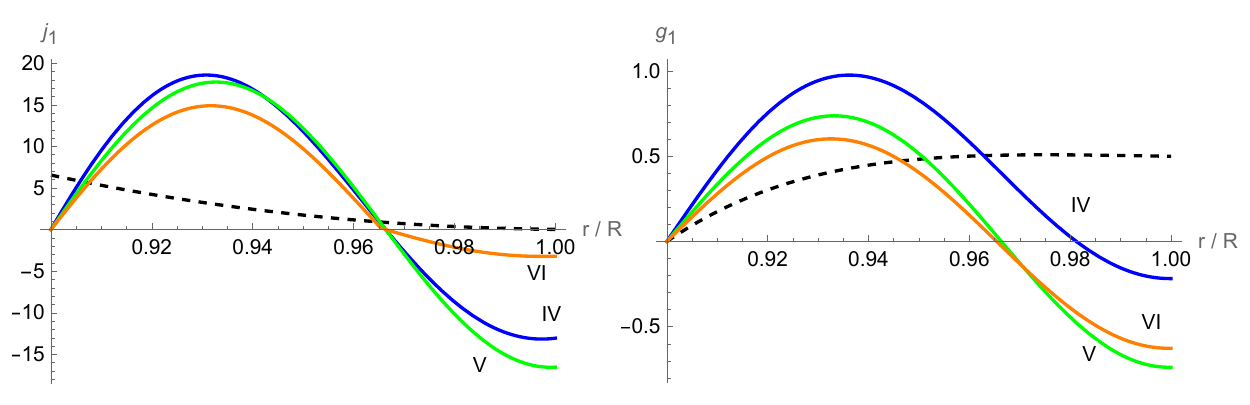}
\caption{ 
 \label{Fig1}
Current distribution $4\pi j_{1} R/(cB_0)$ (left) and solution $g_{1}/(B_0 R^2)$ (right) as a function of $r/R$ for Model I--VI.
Dashed line represents that of Model B.
}
\end{center}\end{figure}

\begin{figure}\begin{center}
  \includegraphics[width=0.35\columnwidth]{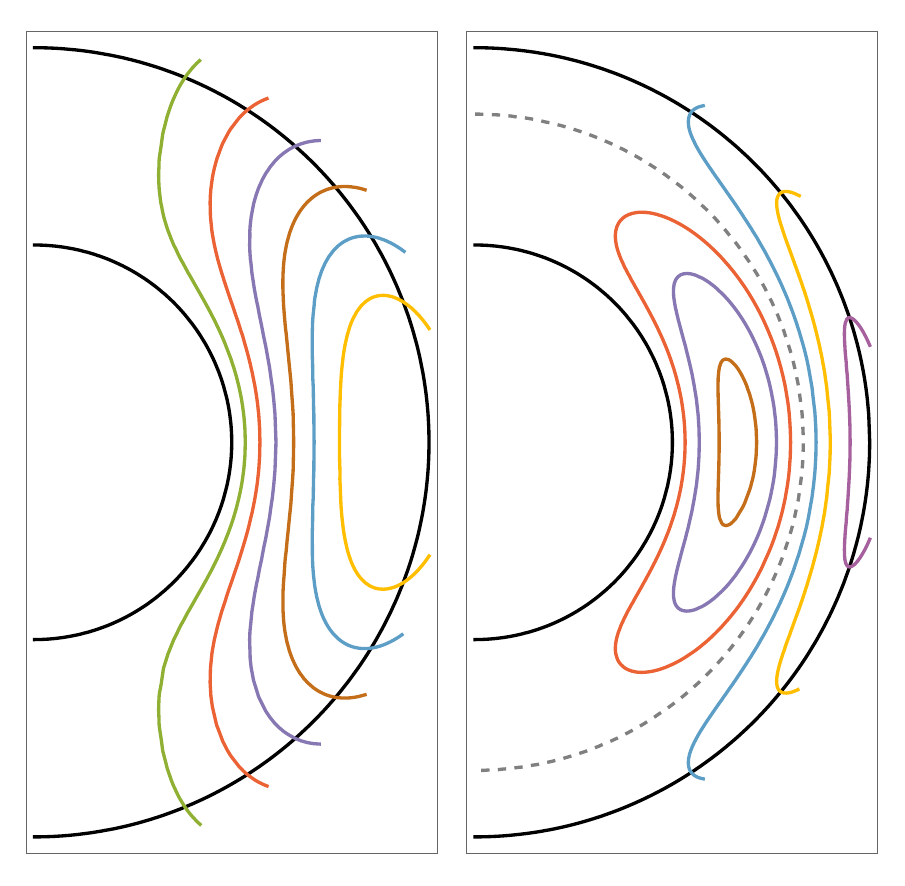}
\caption{ 
 \label{Fig2}
Magnetic field geometry in the meridian plane for Models I—III (left panel) and Models IV—VI (right panel).
Contours of dipolar magnetic function are plotted in the crust enlarged for easy display, a region between two semicircles in solid.
The dashed semicircle on the right panel represents $\Psi=0$, and direction of magnetic field changes across the radius. 
}
\end{center}\end{figure}

%
When a magnetic field is extended inside the core, the strength may be high; hence, the ellipticity of a star is almost determined by the core field. We consider an extreme case in which the field exists outside the core $r>r_c$ to explore the ellipticity due to elastic deformation in the crust\citep[e.g., see Eq.~(19) in][]{2021MNRAS.502.2097K}.
We consider two types of magnetic fields: barotropic and nonbarotropic.
The spatial profiles of the density and pressure are assumed to be nonbarotropic because of the thermal structure at the hot era, denoted as A, and gradually tend to become barotropic as the star cools.
The Lorentz force for a nonbarotropic magnetic field induces rotational and irrotational accelerations. The thermodynamic relation is characterized by $Q \ne 0$ (see Eqs.~(\ref{DefQterm.eqn}) and (\ref{NonB_q.eqn})).
The current distribution $j_{\phi}$ for nonbarotropic magnetic fields is given by the analytic function for Models I--VI in Table~\ref{table:model}.
We also consider a barotropic magnetic field with a constant $k_0$, which induces only irrotational acceleration. 
The magnetic field is similar to the attractor solution arising from the evolution under the Hall effect in the axisymmetric case, because the spatial profile of the electron number density does not differ significantly from that of the mass density~\cite{2014PhRvL.112q1101G,2014MNRAS.438.1618G}.
The magnetic-field configuration is assumed to be realized for the later cold state B, and is referred to Model B.
Equation~(\ref{diffeqg1.eqn}) for the radial function $g_{1}$ is solved with appropriate boundary conditions:
 $g_{1}=0$ at $r_c$ and in vacuum at the surface $R$.
The dipole field strength at the surface for Model B is denoted as $B_{0}$, i.e., $g_1 (R)=B_{0}R^2/2$.
The magnetic field strength in other models is also normalized by $B_{0}$.
In Models IV--VI, the direction of $j_{\phi}$ changes at the middle radius $r/R\approx 0.966$; hence, the direction of the poloidal magnetic field changes from the inner to the outer.
In Model VI, a toroidal magnetic field is considered, confined to the region $r_c< r < 0.966R$.
The radial function $g_1$ is solved using $\kappa^2 g_1$ in the source term~(\ref{sourcej1.eqn}).
The ratio of toroidal to poloidal components in the magnetic energy is $0.456$.  
Figure~\ref{Fig1} shows radial functions $j_{\phi}$ and $g_{1}$ for each model.
Figure~\ref{Fig2} is a schematic of the poloidal magnetic-field geometry in the meridian section. The crustal region was enlarged for ease of display.
The toroidal magnetic field in Model VI is confined to the loop field for $r_c< r < 0.966R$.
The magnetic energy and the total azimuthal current in the crust are denoted as $E_{m}=\int B^2/(8\pi) r^2\sin \theta dr d\theta d\phi$ and $J_{\phi} =\int j_{\phi} rdrd\theta$, respectively.
For Model B, they are denoted as 
$E_{0} \approx 2.02B_{0}^2 R^3$ and $J_{0} \approx 2.42 c B_{0} R$.
Constants $a, b, w$, and $\kappa$ involved in the current distribution of Table~\ref{table:model} are chosen to obtain an appropriate value of $E_m$ and $J_{\phi}$, by which Models I--VI are compared in Table~\ref{table:model}.
We set the energy and total current of the nonbarotropic model to be larger than those of the barotropic model because the former is assumed to change to the latter for some dissipative process.
The total electric current $J_{\phi}$
in Model I is identical to that in Model B. Therefore, the magnetic field in Model I may evolve into that of Model B through a redistribution of the electric current. As shown in the next section, however, the resulting change in ellipticity is small.
To investigate the consequences of the large deformation generated during the early hot stage, we assume a dissipation process in which the electric-current and magnetic energy undergo substantial changes owing to a transition in the thermal structure.
The physical mechanism responsible for such changes remains uncertain and is therefore not specified.
In Models II-VI, the total magnetic energy and electric current significantly differ from those in Model B. Nevertheless, the magnetic field is assumed to evolve toward the same final configuration as that of Model B.
Our concern is the relic of large deformation driven by the magnetic field, which significantly decays.
We cannot follow the evolution, but estimate the residual deformation by comparing equilibrium models.
The perturbations are induced by the Lorentz force; hence, quantities such as $\delta \rho$, $\epsilon$ are proportional to $B_{0}^2$.
We assume $B_{0} \sim 10^{12}-10^{13}$G for a typical pulsar, although $B_{0}$ is mathematically chosen as much larger or smaller values, where different physics should be incorporated.
The typical magnetic energy for $B_{0} =10^{13}$G is $E_{0} \approx 3 \times 10^{44}$ erg. The luminosity is ${\dot{Q}} \approx (10 \times E_{0} -E_{0})/\tau \approx 10^{36}{\rm{erg~s}}^{-1}$, assuming that the magnetic decay from the nonbarotropic state A with $\sim 10 E_0$ to the barotropic state B with $E_0$ in $\tau \approx 10^2$ years after birth.
This luminosity is lower than that for standard cooling~\citep[see, e.g.,][]{2006NuPhA.777..497P,2020ApJ...888...97B}.
Therefore, the transition from state A to B is an allowed scenario, although a drastic change with larger energy would be interesting. 

\section{Results}

\begin{table}[h]
     \caption{Quadrupole ellipticity normalized by $N= 10^{-8}(B_0/10^{13}{\rm{G}})^2$.
    Negative value of $\epsilon$ represents oblate deformation, while positive value represents prolate. Note that the ellipticity sustained by the magnetic field of Model B is $\epsilon_{s}/N$$=-0.7924$. 
    }
    \centering
    \begin{tabular}{c|c|c|c}
        Model & $\epsilon_{a}/N$ &  $\epsilon_{*}/N$ & $\epsilon_{z}/N$\\
         \hline
     I &  $-0.9412$ & $-0.7988$ & $-0.0080$\\
    II & $-5.8584$ & $-0.8158$ & $-0.0249$\\
    III & $-37.305$ & $-0.8588$ & $-0.0680$\\
    IV & $-3.3303$ & $-0.8148$ & $-0.0239$\\
    V &   $+1.8053$ & $-0.8240$ & $-0.0332$\\
    VI &  $+0.2304$ & $-0.7972$ & $-0.0063$\\
    \end{tabular}
\label{table:result}
\end{table}

In Table~\ref{table:result}, numerical results are tabulated for quadrupole ellipticities $\epsilon _a$, $\epsilon _*$, and $\epsilon _z$.
The value $\epsilon _a$, which represents the magneto-hydrodynamic deformation in the hot era, significantly depends on the magnetic geometry and strength.
The deformation is oblate, i.e., $\epsilon _a <0$ for Models I--IV, and prolate, i.e., $\epsilon _a >0$ for Models V and VI.
As shown in Fig.~\ref{Fig1}, the force $\VEC{f}=c^{-1}\VEC{j}\times\VEC{B}$ changes the direction in a certain radial region for Models V and VI.
The deformation is cancelled when calculating the ellipticity through the entire region. The deformed shape is prolate for Model V even for the dipolar magnetic field. The shape of Model VI is also prolate due to the strong toroidal magnetic field. 
The magnitude $|\epsilon _a|$ is of order 
$ 0.5 N (E_{m}/E_{0})$, and increases with $E_{m}/E_{0}$ in Models I--III.
The ellipticity is given by the ratio of magnetic to gravitational energy in the crust:
$|\epsilon | \sim E_m/(G\Delta M M/R)$ $\sim 10^{-8} (B_0/10^{13} {\rm{G}})^2$.
The estimate is appropriate within one order of magnitude despite the various geometries of the magnetic field.
The deformation $\epsilon _a$ was supported by a nonbarotropic magnetic field.
In our evolutionary scenario, the thermal structure tends to a barotropic one as the star cools.
The magnetic field at the later stage is assumed to be given by Model B, in which the magnetic energy is lower than that of the initial nonbarotropic magnetic field.
The ellipticity also changes according to the transition to a new magnetic field. The result $\epsilon _{*}$ is almost the same for all models. 
The final value is approximated as $\epsilon _{*} \approx -0.8\times 10^{-8} (B_0/10^{13} {\rm{G}})^2$, for which the magnetic-field strength $B_0$ at a later time is relevant.
The deformation in the past had little effect on the relic value $\epsilon _{*}$.
The value $\epsilon _{*}$ is approximated by $\epsilon _s$, which is ellipticity sustained by the elasticity and final magnetic field, i.e., that for Model B without the evolutionary effect.
The value $\epsilon _{z}$ is also a relic value when the deforming force
completely vanishes at the later time.
The value $|\epsilon _{z}|$ ranges from $2\times10^{-3}$ to $3\times 10^{-2}$ of $|\epsilon _{a}|$, and is two orders of magnitude smaller than $|\epsilon _{*}|$.
The reduction rate agrees with that reported in a previous study ~\citep{2026MNRAS.546ag141K}, in which the driving force was mathematical.
The reduction rate is explained by the weakness of the elastic force $\mu / p_0 \ll 1$ in the crust, i.e., 
$|\epsilon _{z}|/|\epsilon _{a}|$ $\sim \mu / p_0$  $\sim 10^{-3} - 10^{-2}$, depending on the radius.
The parameter $\zeta$, used to characterize the magnitude of deforming force in the previous study, corresponds to 
$\zeta \sim 10^{-6} (E_{m}/E_0)(B_0/10^{13}{\text{G}})^2$.
The relic $|\epsilon _{z}| \sim 10^{-5}\zeta$ in ~\cite{2026MNRAS.546ag141K}
becomes $|\epsilon _{z}|$ $\sim 10^{-11} (E_{m}/E_0)(B_0/10^{13}{\text{G}})^2$
$=10^{-3} (E_{m}/E_0)N$.
The parameter is constrained by a physical value in the present work.
The Lorentz force at the final state may be neglected when the initial magnetic field is strong enough.
The condition $|\epsilon _{z}| > |\epsilon _{*}|$ corresponds to
$E_{m}> 10^3 E_0=$$3 \times 10^{47} (B_0/10^{13}{\text{G}})^2$ erg and
$B \approx (E_{m}/E_0)^{1/2} B_0>$$3 \times 10^{14} (B_0/10^{13}{\text{G}})$~G.
The initial deformation is large, $|\epsilon _{a}| \sim 10^{-6} (B_0/10^{13}{\text{G}})^2$.
Such a strong magnetic field case would require careful calculations.
%

\begin{figure}\begin{center}
  \includegraphics[width=0.75\columnwidth]{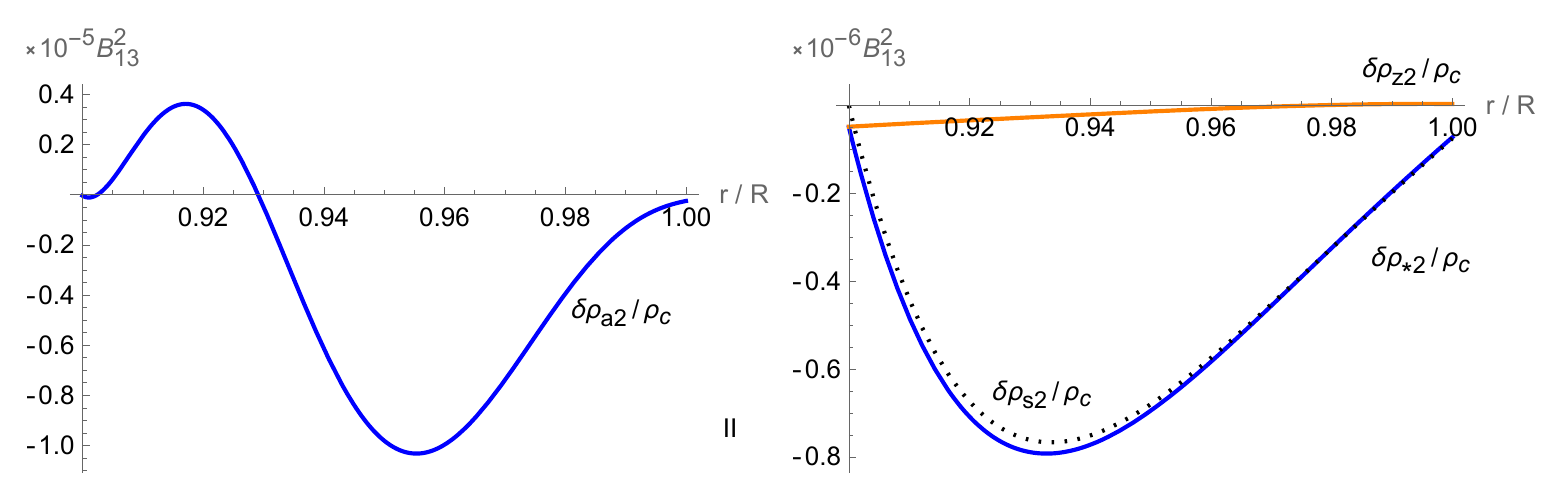}
   \includegraphics[width=0.75\columnwidth]{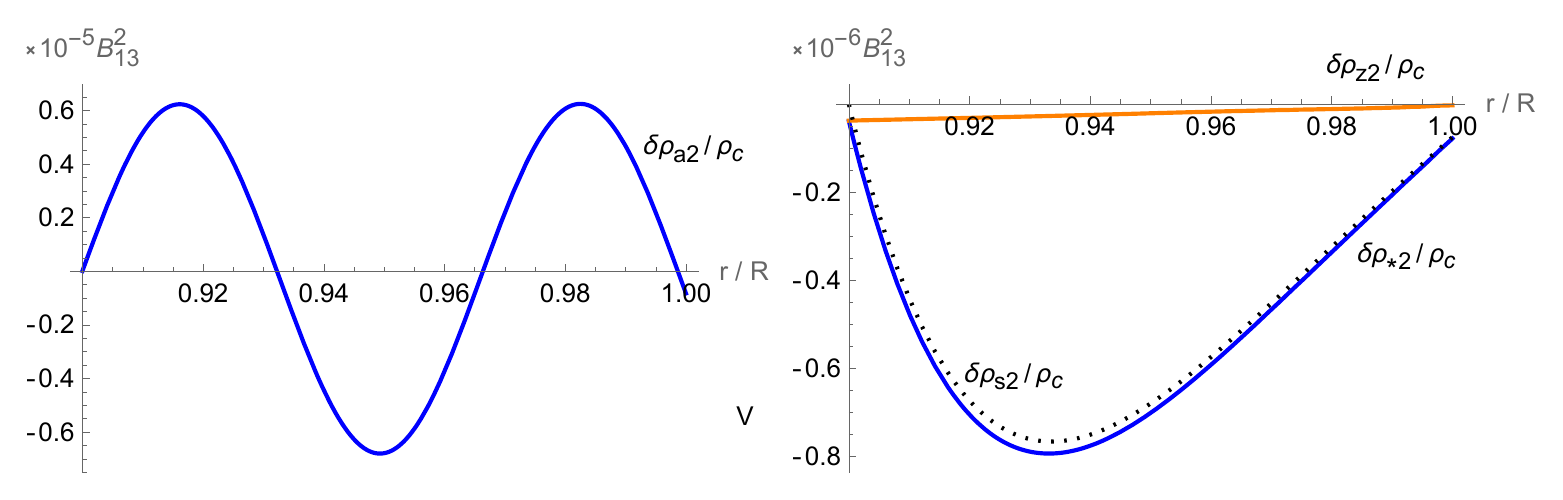}
\caption{ 
 \label{Fig3}
Initial density distribution $\delta \rho_{a2}$ (left) and final one (right) are compared for Models II (top) and V (bottom).
On right panel, radial functions are shown for $\delta \rho_{z2}$ in orange, $\delta \rho_{*2}$ in blue, and $\delta \rho_{s2}$ in black.
}
\end{center}\end{figure}

Figure~\ref{Fig3} shows the radial function $\delta \rho_{a2}$ (left panel) and $\delta \rho_{*2}$ (right panel) in Models II and V.
These functions are the $l=2$ part of the density perturbation relevant to $\epsilon_{a}$ and $\epsilon_{*}$, respectively.
The initial profile depends on the magnetic fields, and $\delta \rho_{a2}$ is different in the two models, leading to different magnitudes of $\epsilon_{a}$.
However, the final density profiles on the right panel are very similar, and slightly depended on the initial condition.
The final profile results from the magnetic field at that time, that of Model B.
On the right panel of Fig.~\ref{Fig3}, the radial function $\delta \rho_{*2}$ is compared with $\delta \rho_{z2}$ and $\delta \rho_{s2}$, which are relevant to $\epsilon_{z}$ and $\epsilon_{s}$, respectively. 
The profile of $\delta \rho_{*2}$ coincides with that of $\delta \rho_{s2}$, but differs from that of $\delta \rho_{z2}$. The magnitude of the latter is significantly small.
The comparison of the radial functions $\delta \rho_{z2}$, $\delta \rho_{*2}$, and $\delta \rho_{s2}$ clearly shows the relationship in the ellipticity $|\epsilon_{z}|$ $\ll |\epsilon_{*}| $  $\sim | \epsilon_{s}|$ in Table \ref{table:result}.

\begin{figure}\begin{center}
 \includegraphics[width=\columnwidth]{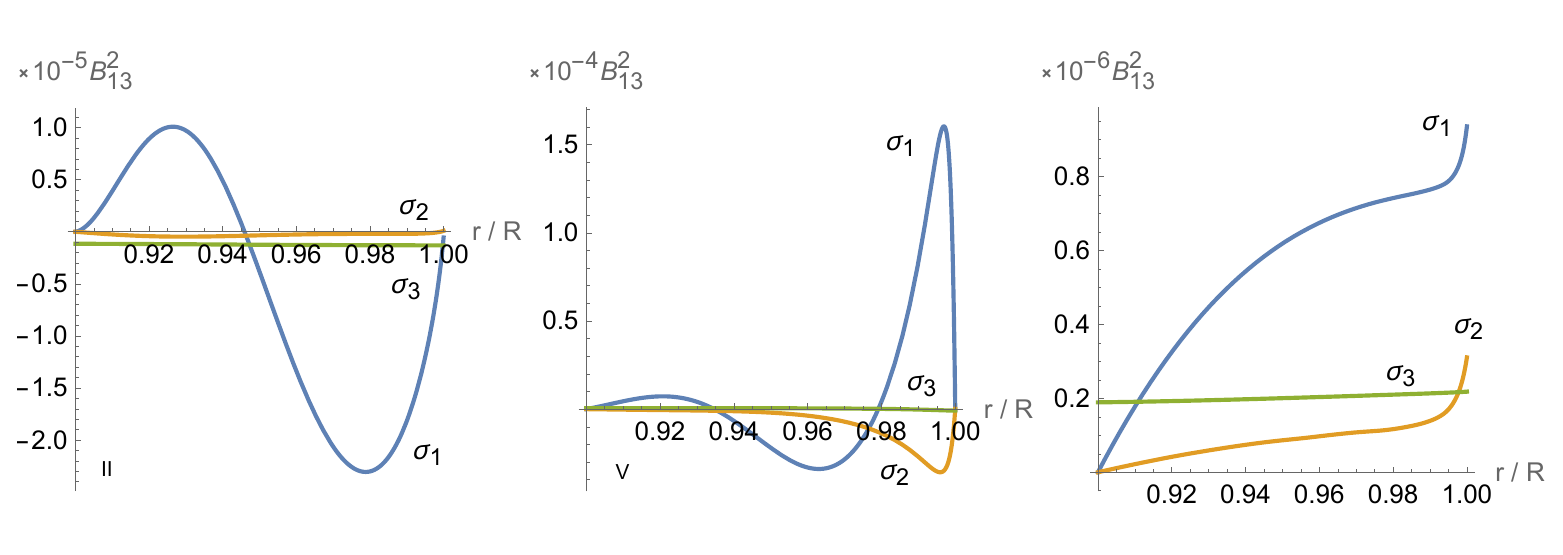}
\caption{ 
 \label{Fig4}
Radial functions $\sigma_{1}, \sigma_{2}$, and $\sigma_{3}$ for shear stress are shown as a function of $r$.
The left and middle panels show results associated with $\epsilon_{*}$
in Models II and V, i.e., residual deformation.
The right panel shows those associated with $\epsilon_{s}$, i.e., magneto-elastic deformation induced by magnetic field Model B.
}

\end{center}\end{figure}

%
Thus, we have understood that $\epsilon_{*}$ is almost determined by the magnetic field at a later time; therefore, $\epsilon_{*}$ $\sim \epsilon_{s}$.
That is, the deformation in the past had little effect on that of the later time.
A question arises.
Is shear stress $\sigma_{ij}$ also the same, irrespective of the initial condition?
Figure~\ref{Fig4} shows three radial functions: $\sigma_{1}, \sigma_{2}$, and $\sigma_{3}$, by which polar components of $l=2$ are expressed as $\sigma_{rr}= \sigma_{1}P_2$, $\sigma_{r\theta}= \sigma_{2}P_{2,\theta}$,
$\sigma_{\theta\theta}=(\sigma_{3} W_{2} -\sigma_{1}P_2)/2$, and $\sigma_{\phi\phi}= -(\sigma_{3} W_{2} +\sigma_{1}P_2)/2$, where $P_2$ is the Legendre polynomial of $l=2$, and $W_{2} =\sin \theta(P_{2,\theta}/\sin\theta)_{,\theta}$.
The shear stress is different in Models II and V, and that of magneto-elastic deformation without an evolutionary scenario.
The magnitude of $\sigma_{1}$, which is the component proportional to $P_2$, is generally large.
The Lorentz force was described by a gradient of the magnetic flux function for the later barotropic state.
The elastic force is a minor component of the force-balance equation.
Therefore, the difference in the shear strain had little effect on the global property of the ellipticity.

%
We also calculated the shear stress magnitude $(\sigma_{ij} \sigma^{ij}/2)^{1/2}$ for Models I--VI.
The maximum was in the range of $\sigma =1-7 \times 10^{-5}(B_0/10^{13}\text{G})^2$.
The deviation associated with the change in magnetic field was small.
The crust was unlikely to fracture because the von Mises criterion was satisfied, i.e., 
$\sigma <$ $\sigma_{c}\approx 10^{-2}-10^{-1}$~\citep{2009PhRvL.102s1102H,2018MNRAS.480.5511B,2018PhRvL.121m2701C}, unless the magnetic field strength $B_0$ exceeds a few times $10^{14}$ G.
%

\section{Summary and Discussion}
We explored the relic of magnetic deformation.
That is, the amount of deformation remaining in the solid crust through an evolutionary scenario, in which a large deformation was assumed to be produced before the solidification at the hot era, and decayed with decreasing Lorentz force.
After the formation of the solid crust, the deformation is imprinted in the crust because the elastic force partially sustains the deformation to cover the decreasing force. 
As a potential method to consider the evolutionary effect, the deformations at two different times were compared.
Our results show that a large deformation in the past was almost unimportant for residual ellipticity.
The equilibrium is governed by the force balance between pressure and gravity, elastic force against the Lorentz forces due to the magnetic field at the later time; however, the contribution of the elastic force was small.
The elastic deformation was so small that the relic remains without fracturing the crust, except for magnetars with $B > 10^{14}$G.
Regardless of the detailed structure with the elastic force, the ellipticity was approximately obtained by the Lorentz force of the magnetic field at the later time.
As discussed in~\cite{2021MNRAS.500.5570G,2022MNRAS.517.5610M}, the residual ellipticity significantly depended on the forces.
The Lorentz force is expressed as $\VEC{f}=[-\Grad B^2/2+(\VEC{B}\cdot\Grad)\VEC{B}]/(4\pi)$, and the gradient term (i.e., magnetic pressure) is important in general.
The change in the magnetic field leads to a small variation in the hydrodynamic pressure, and therefore has little effect on the elastic deformation.
The driving force for deformation at the final state was assumed to be finite, unlike in previous studies.
The final value is approximated as $|\epsilon | \sim 10^{-8} (B_0/10^{13} {\rm{G}})^2$ for magnetic field strength $B_0$ at the surface.
This magnitude is compared with the value $\epsilon \sim 10^{-10} ( B_{\text{av}} /10^{13}{\text{G}})^2$
estimated using the ratio of the magnetic to gravitational energies.
The average value $B_{\text{av}}$ is likely to be larger than $B_{0}$ at the surface because 
the magnetic field in the interior may increase.
For $B_{\text{av}}=10B_{0}$, the ellipticity coincides regardless of the different magnetic field models.
The ellipticity of a superconducting core can be expressed as $\epsilon\sim 3\times 10^{-8} ( B_{0}/10^{13}{\text{G}})( H_{c}/10^{15}{\text{G}})$~\citep{2012MNRAS.419..732L,2014MNRAS.437..424L}.
Moreover, the field configuration significantly depended on the ratio $B/H_c$ at the crust--core boundary; therefore, a detailed model is necessary.
The ellipticity $\epsilon  \sim 10^{-8}$ is commonly expected for the magnetized star with the surface dipole $B_{0} \sim 10^{13}$G, regardless of the crustal elastic force.
Our findings simplify the interpretation of gravitational wave upper limits.
The ellipticity $|\epsilon | \sim 10^{-8} (B_0/10^{13} {\rm{G}})^2$
can be estimated directly from the present-day surface dipole field $B_0$ without requiring detailed knowledge of the proto-NS's magnetic and thermal history. 
This reduces model dependence on constraining neutron star interiors from continuous gravitational wave searches.
In other words, it is difficult to judge which magnetic models are preferable from a limited observation, $\epsilon$ and $B_{0}$, in the current level of theoretical study.
Further studies are required to determine the interior magnetic field.

\section*{Acknowledgments}
This work was supported by JSPS KAKENHI Grant Numbers 
JP23K03389(YK), JP25K17403(AD), and JP26K07066(SK).
\bibliography{kojima26Oct}
\bibliographystyle{aasjournal} 
    \end{document}